\PassOptionsToPackage{dvipsnames}{xcolor}
\documentclass[9pt,sigconf,nonacm, screen]{acmart}

\usepackage{siunitx}
\DeclareSIUnit\cpu{CPU}
\usepackage[all]{nowidow}
\usepackage{xcolor}
\usepackage{subcaption}
\usepackage{hyperref}
\usepackage{acronym}
\usepackage{minted}
\usepackage{tikz}
\usepackage{csquotes}
\usepackage{enumitem}
\usetikzlibrary{tikzmark}

\usepackage[capitalise,noabbrev]{cleveref}
\crefformat{section}{\S#2#1#3} %
\crefformat{subsection}{\S#2#1#3}
\crefformat{subsubsection}{\S#2#1#3}
\crefrangeformat{section}{\S#3#1#4 to~\S#5#2#6}
\crefrangeformat{subsection}{\S#3#1#4 to~\S#5#2#6}
\crefrangeformat{subsubsection}{\S#3#1#4 to~\S#5#2#6}

\usepackage{siunitx}
\DeclareSIUnit\cpu{CPU}
\DeclareSIUnit\vcpu{vCPU}
\begin{document}

\author{Jonas Heisterberg}
\affiliation{%
    \institution{Technische Universit\"at Berlin}
    \city{Berlin}
    \country{Germany}}
\email{jh@3s.tu-berlin.de}
\orcid{0009-0004-4863-4018}

\author{Sebastian Koch}
\affiliation{%
    \institution{Technische Universit\"at Berlin}
    \city{Berlin}
    \country{Germany}}
\email{sk@3s.tu-berlin.de}
\orcid{0009-0003-2296-5305}

\author{Trever Schirmer}
\affiliation{%
    \institution{Technische Universit\"at Berlin}
    \city{Berlin}
    \country{Germany}}
\email{ts@3s.tu-berlin.de}
\orcid{0000-0001-9277-3032}

\author{David Bermbach}
\affiliation{%
    \institution{Technische Universit\"at Berlin}
    \city{Berlin}
    \country{Germany}}
\email{db@3s.tu-berlin.de}
\orcid{0000-0002-7524-3256}

\copyrightyear{2026}
\acmYear{2026}
\setcopyright{cc}
\setcctype{by}
\acmConference[WoAIS '26]{Workshop on AI and Serverless }{December 14--18, 2026}{Tarragona, Spain}
\acmBooktitle{Workshop on AI and Serverless (WoAIS '26), December 14--18, 2026, Tarragona, Spain}
\acmDOI{10.1145/3861043.3861048}
\acmISBN{979-8-4007-3108-2/2026/12}

\newcommand{\name}{AtomicFaaS}
\title{\name{}: Offloading Side-Effect Compensation to the Serverless Platform} %


\begin{abstract}
    Function-as-a-Service (FaaS) simplifies distributed application development by offloading operational concerns to the platform, but developers remain responsible for recovering from failures.
    Retries require idempotent functions, which means that developers have to write complex, error-prone code that undoes or skips the already completed operations of a failed function.

    To address this, we propose \name{}, a platform extension that provides semantic atomicity for the external side effects of a function invocation, whether the function fails or its instance crashes, without developer-written compensation code.
    \name{} intercepts outbound calls and records their effects in a fault-tolerant log, from which the platform compensates leaked effects outside the failed instance. 
    Compared with a manually implemented saga, \name{} eliminates compensation code and leaves no compensatable effects after any tested failure, while adding  \qty{\sim 1}{\milli\second} per intercepted call.
    Under load, it lowers the sustainable throughput of CPU-bound functions, while its latency overhead becomes negligible once calls to external services dominate the invocation time.
\end{abstract}


\maketitle

\section{Introduction}
\label{sec:introduction}
Function-as-a-Service (FaaS) aims to simplify distributed application development by letting developers focus on business logic rather than operational concerns.
The platform handles infrastructure provisioning, resource management, scaling, and fault tolerance~\cite{jonas2019berkeley}.
Developers write applications as small, often stateless functions that the platform invokes in response to events~\cite{hellerstein2018stepback, shahrad2020wild}.

Failure handling, however, has only partly moved to the platform.
It may re-execute a failed invocation, which is safe only for idempotent functions that developers must implement~\cite{jangda2019formal}, and hence the consequences of a failed execution remain the developer's responsibility.
Ensuring idempotency can be challenging~\cite{sreekanti2020aft, ding2023flux}, especially as functions often rely on external services to perform stateful operations~\cite{eismann2021usecases}.
A function may perform several operations with side effects before it fails, so developers must either undo already performed operations or roll forward through them when the function is retried.
While the rollback itself runs within the function instance and cannot protect against failures of that instance, rolling forward requires additional persistence and coordination outside the function for every operation~\cite{zhang2020beldi, jia2021boki}.
Either way, developers have to reason about failure scenarios and write error-prone code that distracts from the business logic, which contradicts the simplicity that FaaS promises.

Following the idea that the system rather than the application should run compensations~\cite{breitbart1993merging, georgakopoulos1995overview}, we introduce \name{}, which brings it to serverless platforms.
Under \name{}, an invocation either commits all of its external side effects, or the platform compensates those that were already applied, independent of the failure cause.
To provide this guarantee, \name{} logs the outbound calls of a function and, if the invocation fails, issues compensating calls from outside the failed instance, using a catalog of inverses maintained by the platform or the service providers.
Once every effect of an invocation has been compensated, it can be retried without producing duplicate effects, and developers write only the simple forward path of their functions.

In this paper, we make the following contributions:
\begin{itemize}
  \item We propose compensating the external effects of failed functions as a platform responsibility and present the design of \name{} (\cref{sec:design}).
  \item We implement an open-source prototype on Knative that requires no changes to Knative or to the function code (\cref{sec:prototype}).
  \item Our evaluation shows that \name{} renders compensation code unnecessary, leaves no compensatable effects behind after any tested failure, and adds \qty{\sim 1}{\milli\second} per intercepted call (\cref{sec:evaluation}).
\end{itemize}

\section{Background \& Related Work}
\label{sec:background}
FaaS abstracts operational tasks away from developers, who write business logic as functions while the platform manages the infrastructure~\cite{jonas2019berkeley}.
Functions are short-lived, stateless, and event-driven~\cite{hellerstein2018stepback, shahrad2020wild}, and the platform treats each invocation as a black box.
This lets it multiplex many functions onto shared resources and scale them independently on demand.
For state and stateful operations, functions rely on external services such as databases, managed services, or third-party APIs~\cite{eismann2021usecases}.
Responsibilities for failure handling are split between the platform and the developers.
For synchronous invocations, every failure, even one caused by the platform, is returned to the client, which has to handle it, most commonly with a retry.
For asynchronous invocations, the platform automates this retry.
Developers, on the other hand, are \emph{required} to write idempotent functions, as functions can fail at any point in their execution and be retried~\cite{jangda2019formal}, a requirement that is challenging to meet~\cite{sreekanti2020aft, ding2023flux}.

\subsection{Atomicity for External Effects}
\label{sec:compensation}
A central concern in distributed systems is that a sequence of operations on different services can fail partway through, leaving partial effects at those services.
Database systems solved this problem decades ago, guaranteeing that a transaction either completes or leaves no lasting change~\cite{gray1981transaction}.
An aborted transaction is undone by restoring the original value of every item it modified, which is possible because the database owns all the data a transaction touches.
In a distributed system, neither the platform nor the application owns the state of external services, so an abort cannot simply restore it~\cite{liskov1983guardians}.
The alternative is compensation, an operation that semantically undoes the effect of a committed operation~\cite{korth1990compensating}.
Sagas build on this idea, a long-running transaction is split into sequential sub-transactions, each paired with a compensating transaction, and when one fails the system runs the compensations of all committed steps in reverse order~\cite{garciamolina1987sagas}.

FaaS workflow engines adopt the same idea, the developer pairs each workflow step with a compensating step and the engine executes them in reverse when a step fails~\cite{cloudflare2026rollbacks, aws2026saga}.
Runtimes for agentic workflows follow, with the compensation for tool calls either provided by a developer~\cite{mohammadi2026atomix} or generated by a model~\cite{chang2025sagallm}.
Compensation can roll back effects on external services, but it weakens atomicity in three ways.
First, effects remain visible to other services until the compensation executes~\cite{garciamolina1987sagas}.
Second, some effects are irreversible, such as a sent e-mail, which cannot be compensated~\cite{liskov1983guardians}.
Third, a compensation restores only a semantically equivalent state, not the original one, and its correctness is not guaranteed~\cite{korth1990compensating}.
Compensation therefore provides semantic atomicity~\cite{garciamolina1983semantic}, not the physical atomicity of a database transaction.
We use the term in this sense throughout the paper: an invocation either commits all of its effects, or every compensatable effect it leaked is semantically undone by the platform.

\subsection{Reliability in FaaS}
\label{sec:related-systems}
Several research systems aim for exactly-once guarantees for serverless functions in the presence of failures~\cite{zhang2020beldi, jia2021boki, qi2023halfmoon, kraft2023apiary}.
They retry a failed function under at-least-once semantics and skip its already completed operations, which gives each operation at-most-once semantics.
While these approaches aim for transactional guarantees on FaaS functions, they do not extend beyond databases, and developers still have to write compensations for any external effects.
Beldi, for instance, is a library that wraps a function's database calls and logs every read, write and invocation, so a retried function skips the operations already done and always rolls forward to completion~\cite{zhang2020beldi}.
It also supports transactions with aborts through shadow tables, but an abort is initiated by the developer and covers only the storage the framework manages.
Apiary instead compiles functions into stored procedures, so that execution is co-located with the data and inherits the DBMS's ACID guarantees~\cite{kraft2023apiary}.
Calls to external services cannot be covered by the database, so Apiary requires them to be idempotent and re-executes them on a retry.
\citeauthor{zhuang2023exoflow}~\cite{zhuang2023exoflow} come closest to our goal, their workflow engine rolls back the external outputs of a failed execution before it replays the workflow, but the developer writes the rollback function for every such output.

Durable execution engines~\cite{burckhardt2021durable}, which commercial platforms have integrated~\cite{aws2026lambdadurable, microsoft2026durablefunctions}, offer the same roll-forward approach as a platform service.
The orchestrator stores the result of each step defined by the developer and, when the function invocation is retried, returns the stored result instead of executing the step again.
Within a step, partial effects remain possible, and the developer has to keep every step idempotent.
None of these systems removes the burden from the developer, who still writes the failure handling, whether as idempotent steps or as compensations.
To the best of our knowledge, no existing platform takes the entire responsibility for undoing the external effects of a failed function.

\begin{figure}
  \centering
  \includegraphics[width=\columnwidth]{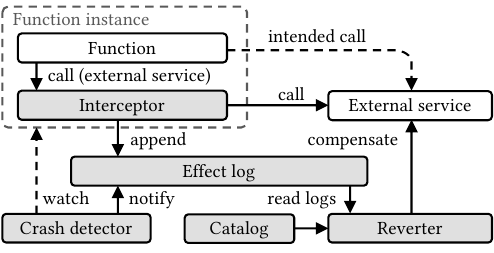}
  \caption{In \name{}, an interceptor logs every outbound call to external services before forwarding it, and a reverter compensates failed invocations according to a shared catalog of inverses.}
  \label{fig:architecture}
\end{figure}

\section{Design}
\label{sec:design}
We propose \name{}, a system that brings semantic atomicity for external effects to FaaS platforms.
An invocation either commits all of its effects or the platform compensates the effects that leaked, even if the function crashes partway through its execution.
The approach requires a platform that can intercept the outbound calls of a function, tell from each response whether the call succeeded, and observe the lifecycle of its function instances.
The unit of this guarantee is the invocation and its side effects, \name{} neither makes a single outbound call atomic nor re-executes the function.

\name{} consists of three core parts that communicate through a shared, low-latency, fault-tolerant effect log, as shown in \cref{fig:architecture}.
The interceptor lives alongside the user code inside the instance and intercepts every outbound call the function instance makes.
Before it forwards a call, it appends a record of the call to the effect log, which lives outside the instance so that it survives an instance crash.
If the function succeeds, the interceptor discards the corresponding entries from the effect log.

An invocation aborts on one of two paths.
On the first, the function returns an error, because an outbound call failed in a way it could not handle or because its own logic rejects the request, and the interceptor marks the invocation as aborted.
On the second, the instance itself crashes, and the crash detector notices the failure from outside the sandbox.
Either way, the reverter is notified, reads the effect log of the aborted invocation, and compensates the recorded calls in reverse order using a catalog of rules.

\subsection{Effect Log and Catalog}
\label{sec:log-catalog}
The interceptor acts as a proxy, appending a record of a call to the effect log before it forwards the call and completing the record with the status and body of the response when it arrives.
Both halves are kept, since the reverter may need values from the response, such as the identifier of a reservation, to construct the neutralizing call.
Because the record precedes the call, a crash leaves the reverter either no record, and then no effect happened, or a record of an effect that may have happened, but never an effect without a record.

The catalog holds rules that recognize a call by its method and path and, if the effect can be undone, describe the call that undoes it, with values taken from the recorded request and response.
The catalog exists once per platform and is shared across all tenants.
We envision the platform owning this catalog, maintaining the rules for its own services, deriving them from API specifications for external ones, and letting third parties contribute rules for their services.

Not every action that leads to a side effect can be compensated, so the catalog also assigns each action one of three classes:
\emph{Invertible} actions have a documented neutralizing operation, such as a payment cancellation or a refund.
Actions with two-phase semantics, such as inventory holds, fall into the same class, their inverse is the release.
The interceptor forwards these calls to the target services, and on abort the reverter executes the neutralizing call.
\emph{Deferrable} actions have an irreversible effect that the function does not need to compute its result, such as sending a confirmation e-mail.
The interceptor does not forward them and immediately returns a placeholder response to the function, which continues its execution but must not depend on the reply.
It forwards them to the target services after the function has returned successfully, a rule adapted from \citeauthor{gray1981transaction}~\cite{gray1981transaction}.
All remaining actions are \emph{uncompensatable}.
The interceptor also stores them in the effect log before forwarding them.
Such an action lies outside the guarantee.
On abort, it is reported for manual resolution or handled by an error handling component, which would also be required if the function did not run on \name{}.

\begin{figure*}[t]
  \centering
  \includegraphics[width=\textwidth]{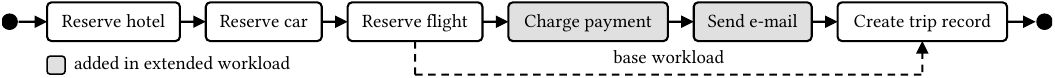}
  \caption{The trip-booking workload~\cite{schmid2025sebsflow} runs in one function, where each step is one outbound call to an external API.}
  \label{fig:trip-workload}
\end{figure*}

\subsection{Commit, Abort, and Compensation}
\label{sec:detection-compensation}
When the function returns, its response passes through the interceptor, which first flushes the deferred actions in the order they were recorded.
It only marks the invocation as committed and answers the invoker once all of them have succeeded.
Hence, a successful response to the caller means that all side effects have occurred, including irreversible ones.
Like any other call, a deferred call can fail, in which case it is retried with an exponential backoff.
If the call ultimately fails, the invocation aborts and any deferred action already delivered is treated as an uncompensatable one.

Whether an invocation commits is decided by the function's response, the flush of its deferred actions, and the survival of its instance.
Beyond a failed flush, the interceptor marks an invocation as aborted if the function returns an error, which also lets a developer abort deliberately without writing any compensation code.
Developers can still decide to handle certain failures in their business logic, so a failed call alone does not abort the invocation.
A crash of the instance, however, takes the interceptor with it, so the platform has to notice it from the outside, which is the job of the crash detector.
Either way, the reverter is notified and processes the recorded actions of an aborted invocation in reverse order.
For each record it looks up the action in the catalog to determine its class and, for an invertible action, instantiates the template of the neutralizing call with values from the recorded request and response and issues it to the service.
Deferred actions that were never forwarded are discarded, uncompensatable actions are reported for manual resolution, and any neutralizing call that still fails after retries is likewise reported.

\begin{figure}
  \centering
  \includegraphics{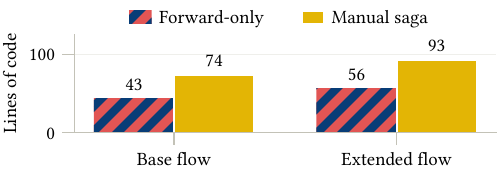}
  \caption{
    Lines of code of the forward-only function on \name{} and the manual saga with its compensation code.}
  \label{fig:e1-loc}
\end{figure}

\section{Evaluation}
\label{sec:evaluation}
We first show that \name{} reverts effects correctly and then quantify what atomicity costs in latency and throughput.

\paragraph{Prototype}
\label{sec:prototype}
We implement \name{} as a proof-of-concept prototype in Go that extends Knative~\cite{knative2026} for HTTP-triggered functions written in JavaScript whose effects stem from outbound HTTP calls.
To leave existing functions unaltered, a shim modifies the Fetch API so that every outbound call is routed through an interceptor running as a proxy next to the function in the same pod.
The interceptor records each call in the effect log, a Redis~\cite{redis2026} store in the same cluster, and then forwards it to the target service.
In production, this store would be replicated for fault tolerance, whereas our prototype runs a single instance.
The crash detector and the reverter also run in the same cluster.
As Knative provides no hook to detect a function failure, the crash detector watches the Kubernetes API for terminated pods and marks the invocations that were running on them as aborted.
The reverter polls the effect log for aborted invocations and issues the inverse of each recorded effect.
We publish the prototype and all results as open source.\!\footnote{\url{https://github.com/persists/atomicfaas}}

\paragraph{Use Case}
We adapt the SeBS-Flow trip-booking workload~\cite{schmid2025sebsflow} and implement it as a single function, which calls a sequence of external APIs, as shown in \cref{fig:trip-workload}.
We refer to the original workload as the base flow and derive an extended flow by adding two effects, a payment, which has to be refunded, and a confirmation e-mail, which is irreversible but deferrable.
Both flows exist in two versions.
The forward-only version issues the calls and nothing else, which is the code a developer would ship on our platform.
The manual saga carries the compensation inside the function, which is what a developer currently has to write for the same outcome.
On a failure, the saga compensates every completed call in reverse order.

\subsection{Developer Effort and Correctness}
First we measure the developer effort \name{} saves, counting the lines of code of both versions with \texttt{scc}~\cite{boyter2026scc}.
As shown in \cref{fig:e1-loc}, the saga grows the base flow from \num{43} to \num{74} lines and the extended flow from \num{56} to \num{93}, an increase of \qty{72}{\percent} and \qty{66}{\percent}, respectively.
Everything the saga adds is compensation, a list of the steps completed so far and a reverse call for each of them, while functions on \name{} are plain without any failure handling.
Compensation code is not difficult to write, but it has to be kept in step with the forward path and tested for every point at which the flow can fail.

Second we show that the compensation works in every tested failure case.
We deploy three versions of the extended flow: the forward-only function on plain Knative, the same function on \name{}, and the manual saga on plain Knative.
We mock the APIs of the external services.
We fail one API at a time along the flow, either by having the mocked API refuse the call or by terminating the instance while a call is in progress.
As expected, the forward-only function on plain Knative leaks every effect completed before the failure.
The manual saga compensates after a refusal but leaks the e-mail once it is sent, as the saga cannot defer it and has no inverse for it.
If the instance is terminated, the saga cannot compensate anything, as its compensation code is gone with the function.
With \name{}, no effect survives the failure in any run, since the e-mail is deferred until the invocation commits and the effects are compensated from outside the function's pod.

\subsection{Cost of Atomicity}
Every request to an external service passes through the interceptor and is recorded before it is forwarded.
To quantify what this costs, we measure the end-to-end latency of a single invocation as well as the latency and throughput under load.

\paragraph{Experimental setup}
The performance experiments run on Google Cloud in zone \texttt{europe-west1}.
Every cluster is a single Kubernetes node running Knative, an \texttt{e2-standard-8} instance with \qty{8}{\vcpu} and \qty{32}{\giga\byte} of memory.
  Every function instance requests \qty{0.15}{\vcpu} and \qty{160}{\mebi\byte} of memory and serves one request at a time.
On the \name{} cluster, the Redis store, the crash detector and the reverter run on the same node and draw from the same resource budget, so the cost of \name{} includes its own infrastructure.
The load generator and the mocked services run on a separate machine to isolate the load from the system under test~\cite{bermbach2017book}.
Every mocked service answers a call after \qty{5}{\milli\second}, including the inverse calls.

\begin{figure}
  \centering
  \includegraphics{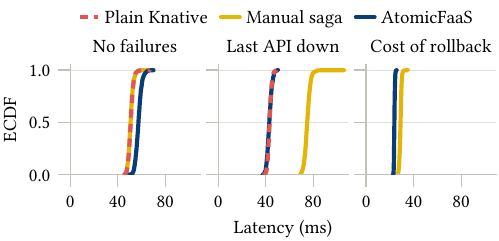}
  \caption{
    End-to-end latency of one invocation without failures and with the last call failing, and rollback duration from the first to the last inverse call at the providers.
    }
  \label{fig:e3-latency}
\end{figure}

\paragraph{Latency of a single invocation}
The first experiment measures the overhead of \name{} when nothing else competes for the node.
We compare the same three versions of the extended flow as before, each deployed on the same machine.
The load generator issues \num{1000} sequential invocations to each version and records the end-to-end latency.
We follow the randomized multiple interleaved trials method, RMIT~\cite{abedi2017rmit}, where a trial is one invocation on one version and each round runs one trial of every version in a fresh random order, to account for autocorrelation between invocations.
\cref{fig:e3-latency} shows the distributions.
Without failures, the median invocation takes \qty{51}{\milli\second} on plain Knative and with the manual saga, and \qty{57}{\milli\second} on \name{}.
\name{} therefore adds \qty{6}{\milli\second} to the six calls of the flow, \qty{\sim 1}{\milli\second} per intercepted call for the write-ahead record and the proxy hop.
When the last call fails, plain Knative and \name{} answer in \qty{43}{\milli\second}, since both return as soon as the refusal arrives.
The manual saga answers in \qty{75}{\milli\second} because it compensates before it replies, while \name{} compensates in the background.
The right panel shows the compensation itself, measured at the mocked services from the first inverse call to the last.
Beyond the \qty{20}{\milli\second} that both spend waiting on the four inverse calls at the mock, the reverter needs \qty{4}{\milli\second} for the compensation and the saga \qty{9}{\milli\second}, so undoing from the platform is not slower than the handwritten code.
Both undo the effects in reverse order one call at a time.
We attribute the difference to the reverter being a compiled service, whereas the saga issues its inverse calls from the interpreted JavaScript runtime of the function.
Overall, \name{} adds \qty{\sim 1}{\milli\second} per intercepted call during normal operation and in contrast to the manual saga, it does not delay the function's response when a failure occurs.

\begin{figure}
  \centering
  \includegraphics{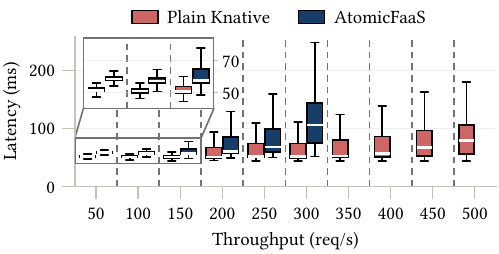}
  \caption{
      Both variants receive increasing request rates while mocked external services answer after \qty{5}{\milli\second}.
    }
  \label{fig:e4-boxes}
\end{figure}

\paragraph{Latency and throughput under load}
The second experiment measures the overhead of \name{} under load, where concurrent invocations share the node.
To this end, we compare the forward-only function on plain Knative and on \name{}, each deployed on a separate \texttt{e2-standard-8} instance.
The load generator drives both deployments in parallel, with one set of mocked API endpoints for each.
Requests are issued in an open loop at a fixed rate, so arrivals do not wait for earlier invocations to finish~\cite{schroeder2006openclosed}.
Every rate runs for \qty{10}{\minute} of measurement after \qty{3}{\minute} of warm-up.
We raise the rate in steps and count a rate as sustained if at least \qty{95}{\percent} of the offered invocations complete successfully.
\cref{fig:e4-boxes} shows the result with the \qty{5}{\milli\second} response delay.
Plain Knative sustains \qty{500}{req\per\second} and \name{} \qty{300}{req\per\second}.
Up to \qty{200}{req\per\second} the median on \name{} stays between \num{57} and \qty{61}{\milli\second} against \qty{51}{\milli\second} on plain Knative, a similar offset as for a single invocation.
At \qty{250}{req\per\second} and \qty{300}{req\per\second} the median rises to \qty{69}{\milli\second} and \qty{106}{\milli\second}, and at \qty{350}{req\per\second} fewer than \qty{95}{\percent} of the invocations complete, as the node is saturated.
Plain Knative reaches the same point at \qty{500}{req\per\second} with a median of \qty{79}{\milli\second}.
The overhead per intercepted call therefore stays roughly constant at moderate load, and saturation arrives earlier for \name{}, which we attribute to the CPU that the interceptors and components consume.

\begin{figure}
  \centering
  \includegraphics{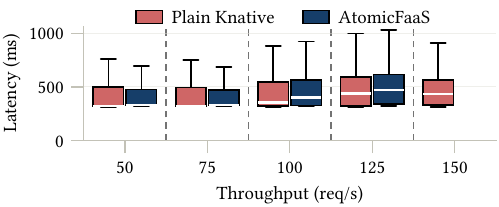}
  \caption{
    Both variants receive increasing request rates while mocked external services answer after \qty{50}{\milli\second}.
    }
  \label{fig:e4-boxes-50ms}
\end{figure}

We repeat the experiment with the mocked API endpoints answering after \qty{50}{\milli\second}, so that an instance spends most of its time waiting for a response.
We also reduce the CPU request of each function instance to \qty{0.075}{\vcpu}, so the node admits more instances and therefore more concurrent waiting invocations, which further increases the load on \name{}.
\cref{fig:e4-boxes-50ms} shows that the difference between the two deployments largely disappears, as the interceptor's work per call becomes a small share of the flow.
Plain Knative sustains \qty{150}{req\per\second} and \name{} \qty{125}{req\per\second}, which we identify as being limited by the number of instances the node admits.
\name{} admits fewer instances because the store, the detector and the reverter draw from the same CPU budget.
None of the components is a scaling bottleneck by design, the interceptor runs in every instance, and the effect log, the detector and the reverter can be partitioned by invocation, but our prototype runs one store and one reverter per cluster, which we did not scale out in this evaluation.

\section{Discussion}
\label{sec:discussion}
We have shown that \name{} reduces developer effort, leaves no compensatable effect behind after any tested failure, where a manual saga still does, and costs \qty{\sim 1}{\milli\second} per intercepted call and a share of the node's resources for its components.
 
\paragraph{Failure handling as a platform responsibility}
Moving failure handling out of the function establishes an additional responsibility for the platform.
The platform takes over a class of failure handling that developers currently write, test and maintain by hand, the same trade the FaaS model already makes for scaling and resource management~\cite{jonas2019berkeley}.
What it costs the developer is a small overhead per intercepted call, which is minor next to the time a call to an external service takes~\cite{ bermbach2020webapibenchmarking2}.
The platform's cost is a share of its resources for the additional components, which it can bill like any other managed service.

Where the failure handling runs matters beyond cost, and it is why \name{} is not a library inside the function, as opposed to other approaches~\cite{zhang2020beldi, jia2021boki, qi2023halfmoon}.
Failure handling inside a function is untrusted tenant code, it runs in the sandbox of one instance, dies with that instance, and knows only what that instance has seen.
Running the effect log and compensation outside this sandbox lets \name{} survive instance crashes.
Compensation inside a function also delays the caller and keeps a fixed-size instance idle on synchronous calls, which the user pays for.
Using \name{}, the platform compensates in the background and answers the caller as soon as the function fails.
\name{} also opens new avenues for future work, as the platform can share aggregate knowledge, such as the failure rates of external services, across functions and tenants to coordinate efficient scheduling and failure handling.
For instance, the platform could postpone invocations that call a failing service or open a circuit breaker for them until the service recovers.

\paragraph{Boundaries of \name{}}
The reverter can only undo what the catalog describes, but the same rollback knowledge is needed today by every developer for every function, whereas \name{} lets the platform acquire it once per service and apply it for all of its clients.
Our prototype intercepts HTTP calls through the patched Fetch API, so effects through other transports escape the interceptor, but the same design extends to other transports the platform can wrap, such as database drivers, queue or gRPC clients.
Lastly, in our prototype, if the instance dies while a call is in flight, the reverter does not know whether the effect exists, and if the inverse needs the provider's response, it cannot issue it and reports the record, which a probe of the service or idempotent operations could address.

\section{Conclusion}
\label{sec:conclusion}
We proposed \name{}, a system that undoes the compensatable external effects of failed function invocations without any developer-written compensation code, which brings FaaS closer to its promise that developers write only their business logic.
It logs every outbound call and compensates leaked effects from outside the failed instance.
Our prototype on Knative spares the \qty{66}{\percent} of additional code that a manual saga adds to an extended trip-booking function and leaves no compensatable effect behind after any tested failure, including instance crashes that a manual saga cannot recover from.
Atomicity costs \qty{\sim 1}{\milli\second} per intercepted call and reduces the sustainable throughput on a CPU-bound workload, while its latency overhead becomes negligible once calls to external services dominate the invocation time.

\begin{acks}
    Funded by \grantsponsor{DFG}{Deutsche Forschungsgemeinschaft (DFG, German Research Foundation)}{https://www.dfg.de/en/} -- \grantnum{DFG}{495343202}.
\end{acks}

\bibliographystyle{ACM-Reference-Format}
\bibliography{bibliography.bib}

\end{document}